\pdfoutput=1
\documentclass[]{ANStopical}
\usepackage{booktabs}
\usepackage{bm}

\newcommand{\nTrainLow}{43}
\newcommand{\nParticlesLow}{100k}
\newcommand{\sigTrainLow}{15.7}
\newcommand{\nTrainMid}{43}
\newcommand{\nParticlesMid}{200k}
\newcommand{\sigTrainMid}{11.3}
\newcommand{\nTrainHi}{43}
\newcommand{\nParticlesHi}{500k}
\newcommand{\sigTrainHi}{7.2}
\newcommand{\nBatches}{600}
\newcommand{\nInactive}{300}
\newcommand{\openmcVersion}{0.15.3}
\newcommand{\boronPpm}{1,360}
\newcommand{\nTrain}{43}
\newcommand{\nDistinct}{40}
\newcommand{\sMin}{2.4}
\newcommand{\sMax}{225.8}
\newcommand{\nTest}{11}
\newcommand{\nLoo}{129}
\newcommand{\nPooledTest}{33}
\newcommand{\nBoot}{20,000}
\newcommand{\nDrawsCheck}{400}

\newcommand{\Tacc}{10}
\newcommand{\sigmaT}{7.25}
\newcommand{\thresh}{12.4}
\newcommand{\totalDirect}{2971}
\newcommand{\totalDirectSd}{10}
\newcommand{\rmseRes}{2.6}
\newcommand{\rmseResPct}{21}
\newcommand{\rmseIntWidthMin}{4}
\newcommand{\rmseIntWidthMax}{13}
\newcommand{\rmseLowFlat}{13.7}
\newcommand{\rmseLoLowFlat}{6.6}
\newcommand{\rmseHiLowFlat}{19.7}
\newcommand{\looRmseLowFlat}{18.8}

\newcommand{\chiMeanLowFlat}{4556}

\newcommand{\rmseLowLogistic}{14.3}
\newcommand{\rmseLoLowLogistic}{8.1}
\newcommand{\rmseHiLowLogistic}{19.0}
\newcommand{\looRmseLowLogistic}{16.6}

\newcommand{\chiMeanLowLogistic}{16.6}

\newcommand{\rmseLowGompertz}{11.1}
\newcommand{\rmseLoLowGompertz}{6.4}
\newcommand{\rmseHiLowGompertz}{15.1}
\newcommand{\looRmseLowGompertz}{16.1}

\newcommand{\chiMeanLowGompertz}{7.7}
\newcommand{\chiJointLowGompertz}{7.3}

\newcommand{\rmseMidFlat}{14.0}
\newcommand{\rmseLoMidFlat}{7.7}
\newcommand{\rmseHiMidFlat}{20.8}
\newcommand{\looRmseMidFlat}{15.9}

\newcommand{\chiMeanMidFlat}{8593}
\newcommand{\chiJointMidFlat}{14.0}

\newcommand{\rmseMidLogistic}{10.3}
\newcommand{\rmseLoMidLogistic}{7.9}
\newcommand{\rmseHiMidLogistic}{12.2}
\newcommand{\looRmseMidLogistic}{17.9}

\newcommand{\chiMeanMidLogistic}{30.3}

\newcommand{\rmseMidGompertz}{10.9}
\newcommand{\rmseLoMidGompertz}{8.0}
\newcommand{\rmseHiMidGompertz}{13.5}
\newcommand{\looRmseMidGompertz}{16.3}

\newcommand{\chiMeanMidGompertz}{13.1}

\newcommand{\rmseHiFlat}{8.8}
\newcommand{\rmseLoHiFlat}{5.2}
\newcommand{\rmseHiHiFlat}{11.8}
\newcommand{\looRmseHiFlat}{12.2}

\newcommand{\chiMeanHiFlat}{21385}

\newcommand{\rmseHiLogistic}{10.6}
\newcommand{\rmseLoHiLogistic}{5.2}
\newcommand{\rmseHiHiLogistic}{14.9}
\newcommand{\looRmseHiLogistic}{10.2}

\newcommand{\chiMeanHiLogistic}{70.2}

\newcommand{\rmseHiGompertz}{9.2}
\newcommand{\rmseLoHiGompertz}{4.8}
\newcommand{\rmseHiHiGompertz}{12.4}
\newcommand{\looRmseHiGompertz}{10.1}

\newcommand{\chiMeanHiGompertz}{29.2}

\newcommand{\phiHiGompertz}{30}
\newcommand{\meanPullHiGompertz}{-0.62}
\newcommand{\meanPullSeHiGompertz}{0.25}
\newcommand{\pairFlatGomHiLo}{-1.42}
\newcommand{\pairFlatGomHiHi}{+0.69}

\newcommand{\totalFlat}{2976}
\newcommand{\totalSdFlat}{17}

\newcommand{\totalDiffFlat}{+5}

\newcommand{\drwPeakFlat}{37.27}
\newcommand{\drwPeakSdFlat}{0.62}
\newcommand{\drwPeakSFlat}{51.0}

\newcommand{\varLooFlat}{1.15}
\newcommand{\varLooLoFlat}{0.91}
\newcommand{\varLooHiFlat}{1.49}
\newcommand{\meanPullLooFlat}{-0.02}
\newcommand{\meanPullLooSeFlat}{0.09}
\newcommand{\covLooFlat}{118}

\newcommand{\varTestFlat}{1.04}
\newcommand{\varTestBlockLoFlat}{0.46}
\newcommand{\varTestBlockHiFlat}{1.49}

\newcommand{\lcRandomFlat}{28}
\newcommand{\lcSpreadFlat}{14}

\newcommand{\totalLogistic}{2989}
\newcommand{\totalSdLogistic}{15}

\newcommand{\totalDiffLogistic}{+18}
\newcommand{\totalDiffSigLogistic}{1.0}
\newcommand{\drwPeakLogistic}{37.32}
\newcommand{\drwPeakSdLogistic}{0.76}
\newcommand{\drwPeakSLogistic}{50.5}

\newcommand{\paramSdPeakLogistic}{0.07}

\newcommand{\varLooLogistic}{1.09}
\newcommand{\varLooLoLogistic}{0.86}
\newcommand{\varLooHiLogistic}{1.41}
\newcommand{\meanPullLooLogistic}{+0.01}
\newcommand{\meanPullLooSeLogistic}{0.09}
\newcommand{\covLooLogistic}{118}

\newcommand{\varTestLogistic}{0.98}
\newcommand{\varTestBlockLoLogistic}{0.47}
\newcommand{\varTestBlockHiLogistic}{1.29}

\newcommand{\totalGompertz}{2982}
\newcommand{\totalSdGompertz}{13}
\newcommand{\totalSdNoCovGompertz}{13.4}
\newcommand{\totalSdCovGompertz}{13.4}
\newcommand{\totalDiffGompertz}{+11}
\newcommand{\totalDiffSigGompertz}{0.7}
\newcommand{\drwPeakGompertz}{37.24}
\newcommand{\drwPeakSdGompertz}{0.78}
\newcommand{\drwPeakSGompertz}{50.5}
\newcommand{\sdEndZeroGompertz}{8.3}
\newcommand{\sdEndTopGompertz}{10.5}
\newcommand{\sdIntMinGompertz}{4.3}
\newcommand{\sdIntMaxGompertz}{5.3}
\newcommand{\paramSdPeakGompertz}{0.08}
\newcommand{\paramSdMaxGompertz}{0.19}
\newcommand{\paramSdMaxSGompertz}{15}
\newcommand{\meanAlonePeakGompertz}{31.5}
\newcommand{\meanAlonePeakSGompertz}{56}
\newcommand{\varLooGompertz}{1.07}
\newcommand{\varLooLoGompertz}{0.85}
\newcommand{\varLooHiGompertz}{1.39}
\newcommand{\meanPullLooGompertz}{+0.01}
\newcommand{\meanPullLooSeGompertz}{0.09}
\newcommand{\covLooGompertz}{119}

\newcommand{\varTestGompertz}{0.90}
\newcommand{\varTestBlockLoGompertz}{0.42}
\newcommand{\varTestBlockHiGompertz}{1.24}

\newcommand{\corrThMinGompertz}{0.38}
\newcommand{\corrThMaxGompertz}{0.54}
\newcommand{\nEffTestGompertz}{17}
\newcommand{\lcRandomGompertz}{20}
\newcommand{\lcSpreadGompertz}{14}

\newcommand{\pairFlatLogLo}{-0.30}
\newcommand{\pairFlatLogHi}{+1.61}
\newcommand{\pairFlatGomLo}{-0.04}
\newcommand{\pairFlatGomHi}{+3.97}
\newcommand{\pairLogGomLo}{+0.16}
\newcommand{\pairLogGomHi}{+2.95}

\newcommand{\kDrwPeakMfive}{37.24}
\newcommand{\kDrwPeakSdMfive}{0.78}

\newcommand{\kDrwSdMedMfive}{0.78}
\newcommand{\kDrwSdEndMfive}{1.7}

\newcommand{\kEllMfive}{24}
\newcommand{\kLmlMfive}{-22.8}
\newcommand{\kLooLpdMfive}{-161.8}

\newcommand{\kVarLooMfiveGom}{1.07}

\newcommand{\kVarLooMfiveFlat}{1.15}

\newcommand{\kTotalMthree}{2984}

\newcommand{\kDrwPeakMthree}{37.40}
\newcommand{\kDrwPeakSdMthree}{1.58}

\newcommand{\kEllMthree}{27}
\newcommand{\kLmlMthree}{-23.8}
\newcommand{\kLooLpdMthree}{-162.4}

\newcommand{\kVarLooMthreeGom}{0.98}

\newcommand{\kVarLooMthreeFlat}{0.93}

\newcommand{\kTotalRbf}{2980}

\newcommand{\kDrwPeakRbf}{37.20}
\newcommand{\kDrwPeakSdRbf}{0.42}

\newcommand{\kEllRbf}{19}
\newcommand{\kLmlRbf}{-20.7}
\newcommand{\kLooLpdRbf}{-159.5}

\newcommand{\kVarLooRbfGom}{1.14}

\newcommand{\kVarLooRbfFlat}{1.36}
\newcommand{\kVarLooRbfFlatLo}{1.08}
\newcommand{\kVarLooRbfFlatHi}{1.76}

\newcommand{\dLmlRbfE}{2.1}
\newcommand{\dLooLpdRbfE}{2.3}

\newcommand{\splineCs}{15.6}
\newcommand{\splineLs}{9.3}
\newcommand{\valNpos}{11}
\newcommand{\valKbiasMean}{8.5}

\newcommand{\valKbiasMin}{-2}
\newcommand{\valKbiasMax}{22}
\newcommand{\valKunc}{7.0}
\newcommand{\valIrwModel}{2984}
\newcommand{\valIrwRef}{2975}
\newcommand{\valIrwDiff}{+8}
\newcommand{\valIrwDiffPct}{+0.28}

\newcommand{\valENDFVIIIBias}{109.3}
\newcommand{\valENDFVIIIBiasUC}{3.0}

\newcommand{\floatnote}[1]{\par\vspace{3pt}\noindent\begin{minipage}{0.94\textwidth}\small #1\end{minipage}\par}

\title{A Gaussian-Process Surrogate for Control Rod Worth with Calibrated Uncertainty: Application to VERA Core Physics Benchmark Problem 4}

\addAuthor{Sean Gohlke}{1}
\addAuthor[jlmiao@psu.edu]{Jilang Miao}{1}
\addAffiliation{1}{Ken and Mary Alice Lindquist Department of Nuclear Engineering, The Pennsylvania State University, University Park, PA 16802, USA}

\Abstract{%
Differential and integral control rod worth are usually obtained by differencing eigenvalue calculations at discrete rod positions,
which resolves the derivative poorly and gives it no uncertainty.
We build a Gaussian-process (GP) surrogate for reactivity as a function of rod position on VERA Core Physics Benchmark Problem~4,
a $3\times3$ assembly cluster with a hybrid AIC/B$_4$C rod cluster control assembly,
modelled in OpenMC with ENDF/B-VIII.0.
A parametric sigmoidal mean carries the dynamic range,
a Mat\'ern-5/2 GP carries the residual,
and the reported Monte Carlo uncertainty of each run enters as known observation noise.
Because differentiation is linear,
the differential rod worth and its variance follow in closed form from the kernel derivatives.
Trained on \nTrain{} runs at three particle counts and tested on \nTest{} independent runs,
the surrogate meets a fixed \Tacc~pcm accuracy target above the test-run noise,
and its intervals are calibrated:
the standardized leave-one-out residuals over \nLoo{} independent runs have variance \varLooGompertz{} with 95\% interval [\varLooLoGompertz,
\varLooHiGompertz].
The total worth agrees with the direct OpenMC end-point difference within combined uncertainty,
and the differential worth peak is located to within one step.
Three mean functions of very different fit quality give the same differential worth to within 0.1~pcm/step,
while the width of its band depends on the kernel smoothness by a factor of four.
The number of runs required and the remaining approximations are reported,
and the extension to rod position and soluble boron is outlined.
}
\keywords{control rod worth, Gaussian process, uncertainty quantification, OpenMC, VERA benchmark}
\shortTitle{Gaussian-Process Surrogate for Control Rod Worth}
\authorHead{S. Gohlke and J. Miao}

\begin{document}

\section{Introduction}\label{sec:intro}

The worth of a control rod bank is a function of its insertion depth.
Two forms of it are used: the integral rod worth (IRW),
the reactivity difference between a given position and a reference position,
and the differential rod worth (DRW), the derivative of reactivity with respect to position.
In continuous-energy Monte Carlo the reactivity at a position is obtained from an eigenvalue calculation carrying a statistical uncertainty of order 10~pcm,
and the usual practice is to run a ladder of positions and difference them.
For the integral worth this is adequate.
For the differential worth,
the difference of two reactivities separated by $\Delta s$ steps estimates a secant rather than a slope,
its statistical error grows as $1/\Delta s$,
and the resulting curve carries no uncertainty.

A surrogate model of reactivity as a smooth function of position addresses both limitations if it can average the Monte Carlo noise of many runs
and attach a tested uncertainty to its own prediction.
Gaussian-process (GP) regression \cite{Rasmussen2006GPML} suits this task for three reasons.
Its posterior is Gaussian, so the prediction at any position comes with a variance.
The known uncertainty of each Monte Carlo run enters the model directly as heteroscedastic observation noise,
so the surrogate uses what the transport code already reports.
And differentiation is a linear operation,
so the derivative of the surrogate is itself a GP whose mean
and variance follow from the derivatives of the kernel.

This paper applies such a surrogate to a single control rod bank in VERA Core Physics Benchmark Problem~4 \cite{Godfrey2014VERA},
modelled in OpenMC \cite{romano2015openmc}.
The contributions are the closed-form integral and differential worth with uncertainty,
a pair of acceptance criteria for accuracy and for the calibration of the uncertainty,
and a study of what the result depends on: the mean function, the kernel,
and the number of training runs.
To our knowledge no GP surrogate for single-bank rod worth with a tested uncertainty statement has been reported at the level of a VERA benchmark problem.
The scope is one input dimension, beginning of life, hot zero power, fixed soluble boron;
the extension to a second input is discussed at the end.

\section{Benchmark Model and Data}\label{sec:data}

\subsection{VERA Problem 4 in OpenMC}\label{sec:model}

VERA Problem~4 is a $3\times3$ array of Westinghouse $17\times17$ assemblies taken from the centre of the Watts Bar Unit~1 initial core,
with a hybrid rod cluster control assembly (AIC tips below B$_4$C) in the centre assembly,
at hot zero power, beginning of life, \boronPpm~ppm soluble boron,
and no thermal-hydraulic feedback \cite{Godfrey2014VERA}.
The rod stroke is 230~steps.
The benchmark specification provides KENO-VI reference eigenvalues at 11 positions from fully inserted to fully withdrawn in 10\% increments,
computed with ENDF/B-VII.0.

The OpenMC (version \openmcVersion) model was built from the benchmark specification
and validated against that reference before the surrogate work began.
With ENDF/B-VII.1, the closest available library to the reference,
the eigenvalue bias over the \valNpos{} reference positions has mean \valKbiasMean~pcm
(range \valKbiasMin{} to \valKbiasMax~pcm,
statistical uncertainty about \valKunc~pcm per run),
and the integral worth from fully inserted to fully withdrawn is \valIrwModel~pcm against the reference \valIrwRef~pcm,
a difference of \valIrwDiff~pcm (\valIrwDiffPct\%).
The surrogate study uses ENDF/B-VIII.0 \cite{Brown2018ENDFVIII},
which shifts the eigenvalue by \valENDFVIIIBias~$\pm$~\valENDFVIIIBiasUC~pcm relative to VII.1 in this model \cite{Kim2021ENDFVII.1/VIII.0}
but leaves the worth,
a difference of eigenvalues, essentially unchanged:
the direct total worth from the VIII.0 test runs described next is \totalDirect~$\pm$~\totalDirectSd~pcm.
\subsection{Training and Test Runs}\label{sec:runs}

The response is reactivity, $\rho = (k-1)/k$ in pcm,
since worth is a reactivity difference and the benchmark tabulates it that way;
the difference between $\Delta k$ and $\Delta\rho$ over the stroke is about 80~pcm,
well above the noise.
Each run reports the standard deviation of $k$ from its batch statistics,
which is converted to $\rho$ through the Jacobian $\sigma_\rho = \sigma_k / k^2$.

The training design is a set of \nTrain{} runs at \nDistinct{} distinct positions between \sMin{}
and \sMax~steps,
space-filling rather than gridded,
with three positions replicated under independent seeds as a check on the reported uncertainty.
The same design was run three times at different particle counts,
so that the effect of label noise on the surrogate can be studied at fixed positions.
Table~\ref{tab:arms} lists the three arms;
every run uses \nBatches{} batches of which \nInactive{} are inactive
and an independent random seed.
Shannon entropy plateaus within about 200 batches at every position except 46 steps,
where it takes about 300, so every run discards 300 batches,
and a split-half comparison of the active batches on the test ladder shows no drift in k.
\begin{table}[htbp]
    \centering
    \caption{Training arms.}
    \begin{tabular}{lccc}
        \toprule
        arm & histories per batch & label noise (pcm) & runs \\
        \midrule
        100k & \nParticlesLow & \sigTrainLow & \nTrainLow \\
        200k & \nParticlesMid & \sigTrainMid & \nTrainMid \\
        500k & \nParticlesHi & \sigTrainHi & \nTrainHi \\
        \bottomrule
    \end{tabular}
    \floatnote{Each arm runs the same \nTrain{} positions with \nBatches{} batches (\nInactive{} inactive) and independent seeds; the label noise is the rms of the reported $\sigma_\rho$.}
    \label{tab:arms}
\end{table}

The test set is \nTest{} separate runs at the benchmark ladder, 0, 23, \ldots, 230 steps,
at 500k histories per batch.
Each of these positions lies between training positions, the nearest 1.2 to 4.3 steps away.
Their rms reported uncertainty is $\sigma_t = \sigmaT$~pcm,
and it enters every accuracy and calibration statement below
because the reference values are themselves measured.
The two ends of the ladder, 0 and 230 steps, lie 2.4 and 4.3 steps outside the training design,
so predictions there are extrapolations.

\section{Method}\label{sec:method}

\subsection{Gaussian-Process Surrogate with Known Noise}\label{sec:gp}

The reactivity at position $s$ is modelled as
\begin{equation}
    \rho(s) = m(s) + f(s), \qquad f \sim \mathcal{GP}\big(0, k(s, s')\big),
    \label{eq:model}
\end{equation}
where $m$ is a parametric mean fitted first by weighted least squares
and $f$ is a zero-mean GP on the residual.
The primary mean is a four-parameter Gompertz curve,
$m(s) = A + B\exp\!\big(-\exp(-(s - s_0)/w)\big)$,
chosen because the reactivity rises steeply as the rod tip rises out of the low-flux region at the bottom of the fuel
and then approaches its asymptote slowly.
A constant mean and a four-parameter logistic are carried through every result as alternatives,
and Section~\ref{sec:sens} shows that all three give the same worth curves.
The mean is a convenience that reduces the dynamic range the GP has to cover,
and its fitted parameters are treated as fixed,
an approximation quantified in Section~\ref{sec:sens}.

The kernel is Mat\'ern-5/2, $k(d) = \sigma_f^2\,(1 + r + r^2/3)\,e^{-r}$ with $r = \sqrt{5}\,|d|/\ell$
and $d = s - s'$,
whose sample paths are twice mean-square differentiable,
so that the derivative process exists and is itself continuous.
The signal variance $\sigma_f^2$
and length scale $\ell$ are set by maximising the marginal likelihood from eight random restarts.
Each training run $i$ contributes an observation $y_i = \rho(x_i) + \varepsilon_i$ with $\varepsilon_i \sim \mathcal{N}(0, \sigma_i^2)$,
where $\sigma_i$ is the reported uncertainty of that run.
With $\mathbf{K}_{ij} = k(x_i, x_j)$, $\mathbf{\Sigma}_n = \mathrm{diag}(\sigma_i^2)$,
$\mathbf{r} = \mathbf{y} - m(\mathbf{x})$
and $\mathbf{k}_*$ the vector of kernel values between a query $s$ and the training positions,
the posterior of the latent reactivity is Gaussian with
\begin{equation}
    \mu(s) = m(s) + \mathbf{k}_*^\top (\mathbf{K} + \mathbf{\Sigma}_n)^{-1} \mathbf{r}, \qquad
    \sigma^2(s) = k(s, s) - \mathbf{k}_*^\top (\mathbf{K} + \mathbf{\Sigma}_n)^{-1} \mathbf{k}_* .
    \label{eq:post}
\end{equation}
The noise enters only through $\mathbf{\Sigma}_n$, as a fixed quantity,
and $\sigma^2(s)$ is the uncertainty about the noise-free reactivity itself;
a further Monte Carlo run adds its own noise on top.
Fifteen independent-seed repeats at each of two positions (0 and 92 steps),
on an earlier revision of the same model,
found the reported $\sigma_k$ to agree with the observed spread within about 15\%,
so the reported values are used as they are.
\subsection{Integral and Differential Worth in Closed Form}\label{sec:drw}

The integral worth is defined here as $\mathrm{IRW}(s) = \rho(s) - \rho(230)$,
the convention of the benchmark's reference table \cite{Godfrey2014VERA},
so that the total worth is $-\mathrm{IRW}(0) = \rho(230) - \rho(0)$.
Because $\rho(s)$ and $\rho(230)$ are jointly Gaussian under the posterior,
$\mathrm{IRW}(s)$ is Gaussian with mean $\mu(s) - \mu(230)$
and variance $\sigma^2(s) + \sigma^2(230) - 2\,\mathrm{Cov}(s, 230)$,
where the posterior covariance is $k(s, 230) - \mathbf{k}_*(s)^\top(\mathbf{K}+\mathbf{\Sigma}_n)^{-1}\mathbf{k}_*(230)$.
The covariance term is retained in the code; in the present design,
with forty training positions between the two ends,
it is numerically zero (total-worth uncertainty \totalSdCovGompertz{} with the term against \totalSdNoCovGompertz~pcm without).

The differential worth is the derivative of Eq.~\eqref{eq:model}.
Differentiation is linear, so $f'$ is a GP whose covariance is $\partial^2 k / \partial s\,\partial s'$,
and the posterior of the derivative at $s$ follows by applying $\partial/\partial s$ to Eq.~\eqref{eq:post}:
\begin{equation}
    \mu'(s) = m'(s) + \mathbf{g}(s)^\top (\mathbf{K} + \mathbf{\Sigma}_n)^{-1} \mathbf{r}, \qquad
    \sigma'^2(s) = k_{11}(0) - \mathbf{g}(s)^\top (\mathbf{K} + \mathbf{\Sigma}_n)^{-1} \mathbf{g}(s),
    \label{eq:drw}
\end{equation}
with $g_i(s) = \partial k(s - x_i)/\partial s$
and $k_{11}(s - s') = \partial^2 k(s, s')/\partial s\,\partial s'$.
For the Mat\'ern-5/2 kernel,
\begin{equation}
    \frac{\partial k}{\partial s} = -\frac{5\sigma_f^2}{3\ell^2}\, d\,(1 + r)\, e^{-r}, \qquad
    k_{11}(d) = \frac{5\sigma_f^2}{3\ell^2}\,\Big(1 + r - \frac{5d^2}{\ell^2}\Big)\, e^{-r},
    \label{eq:kderiv}
\end{equation}
so that $k_{11}(0) = 5\sigma_f^2/(3\ell^2)$ is the prior variance of the slope.
The same expressions were derived for the other kernels used in Section~\ref{sec:sens},
and all are checked against central finite differences at start-up.
The derivative of the fitted mean, $m'(s)$, is exact;
its parameter uncertainty is omitted, and Section~\ref{sec:sens} gives its size.
As a cross-check, \nDrawsCheck{} posterior sample paths differenced numerically give a DRW mean within 0.15~pcm/step of Eq.~\eqref{eq:drw} at the peak,
inside the sampling error of that route.

\subsection{Acceptance Criteria}\label{sec:criteria}

Two criteria are used, one for the posterior mean and one for the posterior standard deviation.

\textit{Accuracy.}
The reference values carry noise $\sigma_t$,
so the measured RMSE at the \nTest{} test positions includes it:
its expectation is $\varepsilon_{\text{surrogate}}^2 + \sigma_t^2$.
The threshold is therefore set above a fixed target $T$ by the test noise:
\begin{equation}
    \mathrm{RMSE} \le \sqrt{T^2 + \sigma_t^2}, \qquad T = \Tacc~\text{pcm}, \quad \sqrt{T^2 + \sigma_t^2} = \thresh~\text{pcm}.
    \label{eq:acc}
\end{equation}
An RMSE over $n$ points has a relative sampling uncertainty of about $1/\sqrt{2n}$,
which for $n = \nTest$ is \rmseResPct\% of its value, or \rmseRes~pcm at the threshold;
every RMSE below therefore carries a bootstrap interval over the test positions \cite{Efron1979Bootstrap},
and models are compared pairwise on the same positions.
Each interval is the 2.5th to 97.5th percentile of the statistic over \nBoot{} resamples of the \nTest{} positions drawn with replacement;
when several fits share the test runs, a position is resampled together with all its arms.

\textit{Calibration.}
For each test run the standardized residual is
\begin{equation}
    z_i = \frac{\mu(x_i) - y_i}{\sqrt{\sigma^2(x_i) + \sigma_i^2}},
    \label{eq:pull}
\end{equation}
with the test run's own noise added to the surrogate variance
because the comparison is against a measured value.
If the surrogate's variance is correct, $z_i \sim \mathcal{N}(0,1)$.
The criterion is that the 95\% chi-square interval on $\mathrm{Var}(z)$ contains 1
(above 1 the bands are too narrow,
below 1 too wide); the coverage of the nominal 95\% interval,
with its Clopper-Pearson interval \cite{ClopperPearson1934}, is the coarser cross-check.
Eleven test points resolve $\mathrm{Var}(z)$ only to within a factor of about three,
so the criterion is applied to the leave-one-out residuals of the training runs,
which are \nLoo{} in number and come from independent seeds.
For a GP the leave-one-out prediction needs no refit:
with $\mathbf{A} = (\mathbf{K}+\mathbf{\Sigma}_n)^{-1}$,
the prediction of $y_i$ from the other runs has mean $y_i - (\mathbf{A}\mathbf{r})_i / A_{ii}$
and variance $1/A_{ii}$ \cite{Sundararajan2001LOO};
that variance already contains $\sigma_i^2$
and is used as the denominator directly.
The hyperparameters are those of the full fit,
which makes the leave-one-out residuals mildly optimistic.

The surrogate is implemented in \texttt{scikit-learn} \cite{Pedregosa2011sklearn},
with the per-run noise supplied as the array of observation variances added to the diagonal of $\mathbf{K}$
and the kernel derivatives,
the closed-form worth and the leave-one-out statistics coded explicitly.

\section{Results}\label{sec:results}

\subsection{Accuracy}\label{sec:acc}

Table~\ref{tab:acc} gives the RMSE at the \nTest{} test positions for each arm
and mean function,
with its bootstrap interval, together with the leave-one-out RMSE over the training runs
and the $\chi^2$ per degree of freedom of the parametric mean fitted alone.
The threshold of Eq.~\eqref{eq:acc} is \thresh~pcm.

\begin{table}[htbp]
    \centering
    \caption{Accuracy at the held-out positions.}
    \small
    \begin{tabular}{llcccc}
        \toprule
        arm & mean & RMSE (pcm) & 95\% interval & LOO RMSE (pcm) & $\chi^2/\mathrm{dof}$, mean alone \\
        \midrule
        100k & constant & \rmseLowFlat & [\rmseLoLowFlat, \rmseHiLowFlat] & \looRmseLowFlat & \chiMeanLowFlat \\
             & logistic & \rmseLowLogistic & [\rmseLoLowLogistic, \rmseHiLowLogistic] & \looRmseLowLogistic & \chiMeanLowLogistic \\
             & Gompertz & \rmseLowGompertz & [\rmseLoLowGompertz, \rmseHiLowGompertz] & \looRmseLowGompertz & \chiMeanLowGompertz \\
        200k & constant & \rmseMidFlat & [\rmseLoMidFlat, \rmseHiMidFlat] & \looRmseMidFlat & \chiMeanMidFlat \\
             & logistic & \rmseMidLogistic & [\rmseLoMidLogistic, \rmseHiMidLogistic] & \looRmseMidLogistic & \chiMeanMidLogistic \\
             & Gompertz & \rmseMidGompertz & [\rmseLoMidGompertz, \rmseHiMidGompertz] & \looRmseMidGompertz & \chiMeanMidGompertz \\
        500k & constant & \rmseHiFlat & [\rmseLoHiFlat, \rmseHiHiFlat] & \looRmseHiFlat & \chiMeanHiFlat \\
             & logistic & \rmseHiLogistic & [\rmseLoHiLogistic, \rmseHiHiLogistic] & \looRmseHiLogistic & \chiMeanHiLogistic \\
             & Gompertz & \rmseHiGompertz & [\rmseLoHiGompertz, \rmseHiHiGompertz] & \looRmseHiGompertz & \chiMeanHiGompertz \\
        \bottomrule
    \end{tabular}
    \floatnote{RMSE at the \nTest{} held-out positions with its 95\% bootstrap interval over positions, leave-one-out (LOO) RMSE over the \nTrain{} training runs, and $\chi^2/\mathrm{dof}$ of the parametric mean fitted alone; Mat\'ern-5/2 kernel; threshold \thresh~pcm.}
    \label{tab:acc}
\end{table}

In the 500k arm every mean function meets the threshold,
with RMSE between
\rmseHiFlat{} %
and
\rmseHiLogistic %
~pcm against the
\sigmaT  %
~pcm rms reported uncertainty of the test runs;
the surrogate averages the
\sigTrainHi~pcm %
rms reported uncertainty of its own training runs,
at the same particle count,
down by a factor of about 1.5 at the interior positions,
where the posterior standard deviation is \sdIntMinGompertz{} to \sdIntMaxGompertz~pcm.
In the 200k arm the parametric means meet the threshold and the constant mean does not;
in the 100k arm only the Gompertz mean does;
but these distinctions are at the resolution of the test.
The bootstrap intervals in Table~\ref{tab:acc} span \rmseIntWidthMin{} to \rmseIntWidthMax~pcm,
and the paired comparison of the three means on the same positions,
pooled over the three arms
and resampling the eleven positions as blocks
so that the shared test runs are handled correctly,
gives 95\% intervals on the RMSE difference of [\pairFlatLogLo,
\pairFlatLogHi]~pcm for constant against logistic, [\pairFlatGomLo,
\pairFlatGomHi]~pcm for constant against Gompertz, and [\pairLogGomLo,
\pairLogGomHi]~pcm for logistic against Gompertz;
in the 500k arm alone the constant-against-Gompertz interval is [\pairFlatGomHiLo,
\pairFlatGomHiHi]~pcm.
The first two contain zero;
the third suggests that Gompertz is the more accurate of the two parametric means,
by less than the \rmseRes~pcm resolution of a single RMSE.
At these sample sizes the three means are therefore equally accurate within the resolution,
and every mean alone leaves structure at the noise level:
its $\chi^2/\mathrm{dof}$ is far above one,
so the GP residual carries the shape.

As a classical reference on the 500k arm,
a natural cubic spline through the training runs gives an RMSE of \splineCs~pcm at the test positions
and a weighted least-squares cubic spline with six interior knots gives \splineLs~pcm,
against \rmseHiGompertz~pcm for the GP with the Gompertz mean.
On accuracy alone the smoothing spline is as good as the GP;
the remaining sections test what the GP adds,
a prediction interval and a derivative with a band.

\subsection{Calibration}\label{sec:cal}

\begin{figure}[htbp]
    \centering
    \includegraphics{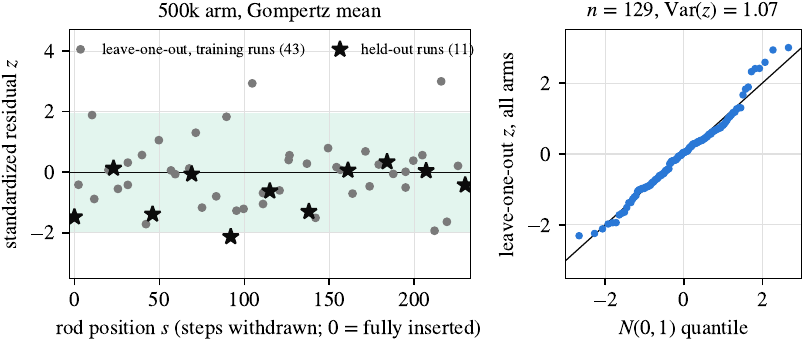}
    \caption{Standardized residuals.}
    \floatnote{Left: residuals Eq.~\eqref{eq:pull} of the 500k arm with the Gompertz mean, leave-one-out for the \nTrain{} training runs and held-out for the \nTest{} test runs; the shaded band is $\pm1.96$. Right: normal quantile plot of the \nLoo{} pooled leave-one-out residuals over the three arms.}
    \label{fig:cal}
\end{figure}

Figure~\ref{fig:cal} shows the standardized residuals of the 500k arm,
both leave-one-out and held-out,
and the normal quantile plot of the pooled leave-one-out residuals over all three arms.
Table~\ref{tab:cal} gives the calibration statistics for each mean function.

\begin{table}[htbp]
    \centering
    \caption{Calibration by mean function.}
    \small
    \begin{tabular}{lcccccc}
        \toprule
        & \multicolumn{4}{c}{leave-one-out, $n = \nLoo$} & \multicolumn{2}{c}{held-out, $n = \nPooledTest$} \\
        \cmidrule(lr){2-5} \cmidrule(lr){6-7}
        mean & Var($z$) & 95\% interval & mean $z$ & coverage & Var($z$) & block-bootstrap interval \\
        \midrule
        constant & \varLooFlat & [\varLooLoFlat, \varLooHiFlat] & \meanPullLooFlat{} $\pm$ \meanPullLooSeFlat & \covLooFlat/\nLoo & \varTestFlat & [\varTestBlockLoFlat, \varTestBlockHiFlat] \\
        logistic & \varLooLogistic & [\varLooLoLogistic, \varLooHiLogistic] & \meanPullLooLogistic{} $\pm$ \meanPullLooSeLogistic & \covLooLogistic/\nLoo & \varTestLogistic & [\varTestBlockLoLogistic, \varTestBlockHiLogistic] \\
        Gompertz & \varLooGompertz & [\varLooLoGompertz, \varLooHiGompertz] & \meanPullLooGompertz{} $\pm$ \meanPullLooSeGompertz & \covLooGompertz/\nLoo & \varTestGompertz & [\varTestBlockLoGompertz, \varTestBlockHiGompertz] \\
        \bottomrule
    \end{tabular}
    \floatnote{Variance of the standardized residuals with its 95\% chi-square interval, mean residual, and coverage of the nominal 95\% interval, for the pooled leave-one-out residuals ($n = \nLoo$, independent seeds) and for the held-out pool ($n = \nPooledTest$, correlated; see text). Mat\'ern-5/2 kernel.}
    \label{tab:cal}
\end{table}

For all three means the leave-one-out variance is consistent with one
and the mean residual with zero,
so the criterion is met, and the quantile plot shows no departure from normality beyond the two
or three largest residuals expected in a sample of \nLoo.
The intervals are about $\pm20$\%, which is the resolution that \nLoo{} points afford;
calibration separates the three means as little as accuracy does.
In the 500k arm the held-out residuals share an offset, visible in Figure~\ref{fig:cal}:
their mean is \meanPullHiGompertz~$\pm$~\meanPullSeHiGompertz{} with the Gompertz mean
and similar with the other two.
It is one arm of three, it is inside the joint test below,
and the leave-one-out pool shows no offset.

The held-out pool is correlated:
the same \nTest{} test runs are scored against the three arms,
so the test-run noise is common to the three residuals at each position.
The surrogate errors are independent across arms (independent training seeds),
and the correlation between two arms' standardized residuals at a position is $\sigma_i^2/\sqrt{(\sigma_a^2+\sigma_i^2)(\sigma_b^2+\sigma_i^2)}$,
which evaluates to \corrThMinGompertz{} to \corrThMaxGompertz{} here.
A pool of \nPooledTest{} such residuals has an effective size near \nEffTestGompertz,
so a chi-square interval computed as if they were independent is too narrow.
The intervals in the last column of Table~\ref{tab:cal} are therefore from a bootstrap that resamples positions as blocks.
They contain one, but they are wide.

Two further checks were made.
First, the standardized residuals test only the diagonal of the posterior covariance,
whereas the worth quantities inherit its off-diagonal structure;
the Mahalanobis statistic of the \nTest{} held-out residuals against the full predictive covariance,
$\mathbf{r}^\top(\mathbf{\Sigma}_* + \mathrm{diag}\,\sigma_t^2)^{-1}\mathbf{r}$,
is $\chi^2_{11}$ under calibration
(95\% range 3.8 to 21.9) and lies between \chiJointLowGompertz{}
and \chiJointMidFlat{} over all arms and means.
Second, the held-out test has limited power in the 500k arm
because the test-run noise is the larger part of the predictive variance:
the surrogate's share at the interior test positions is \phiHiGompertz\%.
A surrogate band twice too narrow would move the held-out Var($z$) only from 1 to about 1.3,
which is inside the resolution of \nTest{} points;
a sharper direct test would need reference runs with lower noise.

\subsection{Worth Curves}\label{sec:worth}

\begin{figure}[htbp]
    \centering
    \includegraphics{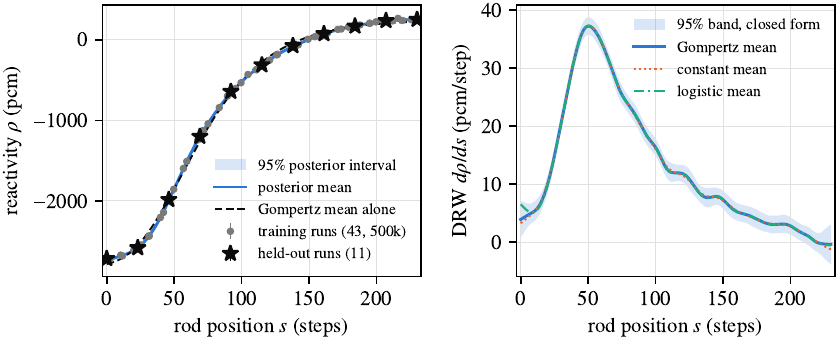}
    \caption{Reactivity and differential rod worth, 500k arm.}
    \floatnote{Gompertz mean, Mat\'ern-5/2 kernel. Left: posterior mean and 95\% interval of the reactivity, the fitted mean alone, and the training and held-out runs (error bars $\pm1.96\sigma$, mostly smaller than the markers). Right: differential rod worth from Eq.~\eqref{eq:drw} with its 95\% band; the DRW means obtained with the constant and logistic means coincide with it.}
    \label{fig:worth}
\end{figure}

Figure~\ref{fig:worth} shows the surrogate for the 500k arm with the Gompertz mean:
the reactivity with its 95\% posterior interval, the training and test runs,
and the differential worth from Eq.~\eqref{eq:drw} with its closed-form band.
The total worth from the surrogate is \totalGompertz~$\pm$~\totalSdGompertz~pcm,
against \totalDirect~$\pm$~\totalDirectSd~pcm from the direct difference of the two end-point test runs,
a difference of \totalDiffGompertz~pcm or \totalDiffSigGompertz{} combined standard deviations.
The differential worth peaks at \drwPeakGompertz~$\pm$~\drwPeakSdGompertz~pcm/step at $s = \drwPeakSGompertz$~steps,
with the rod tip in the lower part of the active fuel.
The band on the derivative is between \kDrwSdMedMfive{}
(median over the stroke) and \kDrwSdEndMfive~pcm/step at the two ends;
the end values are larger because both ends lie outside the training design.
The posterior standard deviation of the reactivity shows the same extrapolation,
\sdEndZeroGompertz{} and \sdEndTopGompertz~pcm at 0 and 230 steps against \sdIntMinGompertz{} to \sdIntMaxGompertz~pcm inside,
and the total worth, evaluated at the two ends, carries that extra uncertainty;
two training runs at the end positions would remove it.

\subsection{Sensitivity to the Mean Function and the Kernel}\label{sec:sens}

Table~\ref{tab:mean} collects the worth quantities for the three mean functions.
The means differ greatly as standalone fits:
their $\chi^2/\mathrm{dof}$ ranges from \chiMeanHiGompertz{} to \chiMeanHiFlat,
and the Gompertz curve alone puts the DRW peak at \meanAlonePeakGompertz~pcm/step at $s = \meanAlonePeakSGompertz$,
some 16\% below and five steps beyond the full model's.
Yet the full-model DRW peaks agree to 0.1~pcm/step, a tenth of their bands,
and the three total worths lie within \totalDiffSigLogistic{} combined standard deviations of the direct value.
The GP residual supplies the shape the mean lacks in each case,
and the derived quantities are the same whichever mean was subtracted.

\begin{table}[htbp]
    \centering
    \caption{Worth quantities by mean function, 500k arm.}
    \small
    \begin{tabular}{lccccc}
        \toprule
        mean & total worth & minus direct & DRW peak & peak position & parameter term \\
             & (pcm) & (pcm) & (pcm/step) & (steps) & (pcm/step) \\
        \midrule
        constant & \totalFlat{} $\pm$ \totalSdFlat & \totalDiffFlat & \drwPeakFlat{} $\pm$ \drwPeakSdFlat & \drwPeakSFlat & 0 \\
        logistic & \totalLogistic{} $\pm$ \totalSdLogistic & \totalDiffLogistic & \drwPeakLogistic{} $\pm$ \drwPeakSdLogistic & \drwPeakSLogistic & \paramSdPeakLogistic \\
        Gompertz & \totalGompertz{} $\pm$ \totalSdGompertz & \totalDiffGompertz & \drwPeakGompertz{} $\pm$ \drwPeakSdGompertz & \drwPeakSGompertz & \paramSdPeakGompertz \\
        \bottomrule
    \end{tabular}
    \floatnote{Mat\'ern-5/2 kernel. The parameter term is the delta-method contribution of the fitted mean's parameter covariance to the DRW standard deviation at the peak, which the closed-form band omits.}
    \label{tab:mean}
\end{table}

The one approximation attached to a parametric mean is that its fitted parameters are treated as exact when the posterior is differentiated.
The omitted term is bounded by propagating the least-squares parameter covariance through $m'(s)$ with the delta method
(conservative when added in quadrature,
since the sequential fit anticorrelates the two errors).
For the Gompertz mean it is \paramSdPeakGompertz~pcm/step at the DRW peak against the \drwPeakSdGompertz~pcm/step of the GP band,
and at most \paramSdMaxGompertz~pcm/step anywhere on the stroke
(at $s = \paramSdMaxSGompertz$).
The constant mean has no such term, and its band is the whole answer;
for the parametric means the omission is quantified here and is a tenth of the band.

\begin{figure}[htbp]
    \centering
    \includegraphics{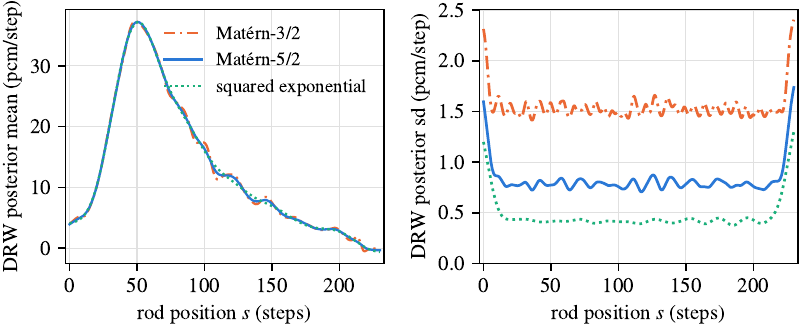}
    \caption{Kernel dependence of the differential worth and its band.}
    \floatnote{500k arm, Gompertz mean. Left: posterior mean of the differential worth for three kernels. Right: its posterior standard deviation from Eq.~\eqref{eq:drw}, which is set by the smoothness of the kernel rather than by the data.}
    \label{fig:kernel}
\end{figure}

The kernel is a different matter.
Figure~\ref{fig:kernel} and Table~\ref{tab:kernel} compare the Mat\'ern-3/2,
Mat\'ern-5/2 and squared-exponential kernels with the Gompertz mean on training-only criteria
(log marginal likelihood and leave-one-out log predictive density),
the held-out RMSE for reference, and the DRW band.
A rational-quadratic kernel was also fitted; its shape parameter ran to its upper bound,
where the kernel is the squared exponential, so it duplicates that result.

\begin{table}[htbp]
    \centering
    \caption{Kernel study, 500k arm.}
    \small
    \begin{tabular}{lcccccc}
        \toprule
        kernel & $\ell$ (steps) & LML & LOO lpd & DRW peak (pcm/step) & Var($z$) Gompertz & Var($z$) constant \\
        \midrule
        Mat\'ern-3/2 & \kEllMthree & \kLmlMthree & \kLooLpdMthree & \kDrwPeakMthree{} $\pm$ \kDrwPeakSdMthree & \kVarLooMthreeGom & \kVarLooMthreeFlat \\
        Mat\'ern-5/2 & \kEllMfive & \kLmlMfive & \kLooLpdMfive & \kDrwPeakMfive{} $\pm$ \kDrwPeakSdMfive & \kVarLooMfiveGom & \kVarLooMfiveFlat \\
        squared exp. & \kEllRbf & \kLmlRbf & \kLooLpdRbf & \kDrwPeakRbf{} $\pm$ \kDrwPeakSdRbf & \kVarLooRbfGom & \kVarLooRbfFlat \\
        \bottomrule
    \end{tabular}
    \floatnote{Gompertz mean. Log marginal likelihood (LML) and leave-one-out log predictive density (LOO lpd) are training-only criteria. The DRW standard deviation is at the peak; the last two columns are the pooled leave-one-out Var($z$) over the three arms for the Gompertz and constant means (95\% intervals are about $\pm20$\%).}
    \label{tab:kernel}
\end{table}

The posterior mean of the differential worth is the same for all three kernels,
with peaks between \kDrwPeakRbf{} and \kDrwPeakMthree~pcm/step,
and so is the total worth (\kTotalRbf{} to \kTotalMthree~pcm).
The band differs: its standard deviation at the peak is \kDrwPeakSdRbf,
\kDrwPeakSdMfive{} and \kDrwPeakSdMthree~pcm/step for the squared-exponential,
Mat\'ern-5/2 and Mat\'ern-3/2 kernels, a factor of four between the smoothest and the roughest.
The training data prefer the squared exponential,
by \dLmlRbfE{} in log marginal likelihood
and \dLooLpdRbfE{} in leave-one-out log predictive density on the 500k arm
and by less on the others,
which is weak evidence;
43 points leave the smoothness of the process open.
The width of the derivative band is therefore a modelling choice, made here a priori:
the Mat\'ern-5/2 kernel gives a derivative that is once mean-square differentiable,
and its band lies between the other two.
The narrower squared-exponential band has a cost: with the constant mean,
whose GP has to carry the entire shape,
the squared-exponential kernel fails the leave-one-out calibration criterion
(Var($z$) = \kVarLooRbfFlat,
interval [\kVarLooRbfFlatLo, \kVarLooRbfFlatHi]),
whereas the Mat\'ern kernels pass with every mean.
We report the Mat\'ern-5/2 band and give the other two as its range.

\subsection{Number of Training Runs}\label{sec:volume}

\begin{figure}[htbp]
    \centering
    \includegraphics{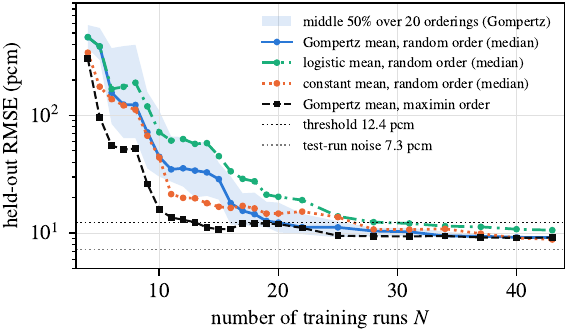}
    \caption{Held-out RMSE against the number of training runs, 500k arm.}
    \floatnote{Median over 20 random orderings for each mean (middle 50\% shaded for Gompertz), and the maximin order for the Gompertz mean.}
    \label{fig:learn}
\end{figure}

Figure~\ref{fig:learn} shows the held-out RMSE of the 500k arm as training runs are added,
for the three mean functions in random order
(median over 20 orderings) and for the Gompertz mean with runs added in a maximin order that spreads them over the stroke.
The threshold Eq.~\eqref{eq:acc} is reached
and held from \lcRandomGompertz{} runs in random order
and \lcSpreadGompertz{} runs in the spread order with the Gompertz mean,
and from \lcRandomFlat{} and \lcSpreadFlat{} with the constant mean.
The crossing points carry several runs of uncertainty (Section~\ref{sec:criteria});
the firmer statement is that a spread design reaches the target with about two thirds of the runs a random one needs,
and that the parametric mean helps most at small $N$,
where the GP has little to learn the shape from.

\section{Conclusions}\label{sec:concl}

A Gaussian-process surrogate with a parametric mean,
a Mat\'ern-5/2 kernel and the reported Monte Carlo uncertainty of each run as known observation noise reproduces the reactivity of a single control rod bank in VERA Problem~4 to within a fixed \Tacc~pcm target above the test-run noise,
from \nTrain{} OpenMC runs at 500k histories per batch, and its uncertainty is calibrated:
the standardized leave-one-out residuals over \nLoo{} independent runs have variance \varLooGompertz{} with interval [\varLooLoGompertz,
\varLooHiGompertz].
The differential rod worth and its band follow in closed form from the kernel derivatives,
agree with a sampled cross-check,
and locate the peak at \drwPeakGompertz~$\pm$~\drwPeakSdGompertz~pcm/step at \drwPeakSGompertz~steps;
the total worth agrees with the direct end-point difference within \totalDiffSigGompertz{} combined standard deviations.
The results are the same for the three mean functions:
three means of very different standalone fit quality give the same worth curves,
and the omitted parameter uncertainty of the fitted mean is a tenth of the derivative band.
The kernel does matter to the band:
its width varies by a factor of four across the smoothness classes examined,
the data prefer the smoothest only weakly,
and the smoothest is overconfident when the GP carries the whole shape,
so the band reported is conditional on a smoothness choice made in advance
and the range across kernels is part of the result.

Three limitations remain.
The kernel hyperparameters are point estimates, so the band omits their uncertainty.
The two ends of the stroke lie outside the training design,
and the total worth is evaluated there; two additional runs would remove that extrapolation.
And the criteria were evaluated on one benchmark problem and one bank,
so the data-volume conclusion, about 20 well-spread runs at 500k histories,
applies to a response of this shape and noise level.
The next steps are to combine the three arms, which share their positions,
by inverse-variance weighting,
which lowers the effective label noise below that of the best single arm,
and to extend the input space to rod position and soluble boron with a two-dimensional kernel.

\section*{ACKNOWLEDGEMENTS}
The authors thank the Pennsylvania State University Institute for Computational
and Data Sciences (ICDS) for computational resources support.

\bibliographystyle{ANStopical}
\bibliography{references}

\end{document}